%                                  G. 't Hooft  macros version Oct.96
\newread\testifexists
\def\GetIfExists #1 {\immediate\openin\testifexists=#1
	\ifeof\testifexists\immediate\closein\testifexists\else
        \immediate\closein\testifexists\input #1\fi}
\def\epsffile#1{Figure: #1} 	%%%

\immediate\openin\testifexists=e
	\ifeof\testifexists\immediate\closein\testifexists\else
        \immediate\closein\testifexists\input e\fipsf 
  
\magnification= \magstep1	% or use \magstephalf
\tolerance=1600 
\parskip=5pt 
\baselineskip= 5 true mm \mathsurround=1pt
\font\smallrm=cmr8

\font\medrm=cmr9

\font\bigbf=cmbx12
 	\def\Bbb#1{\setbox0=\hbox{$\tt #1$}  \copy0\kern-\wd0\kern .1em\copy0} 
	\immediate\openin\testifexists=a
	\ifeof\testifexists\immediate\closein\testifexists\else
        \immediate\closein\testifexists\input a\fimssym.def %% for \Bbb A - Z %%
\def\secbreak{\vskip12pt plus .3in \penalty-60\vskip 0pt plus -.2in} 
 %\prefbreak{distance}
\def\hugeskip{\vskip12mm plus 3mm}
\def\Narrower{\par\narrower\noindent}	% never again use TeX's awful \narrower
\def\Endnarrower{\par\leftskip=0pt \rightskip=0pt} 
	\def\ra{\rightarrow}		
\def\a{\alpha}          \def\b{\beta}   \def\g{\gamma}  
\def\d{\delta}          \def\D{\Delta}  \def\e{\varepsilon}
              \def\l{\lambda}          
\def\m{\mu}             \def\f{\phi}    \def\F{\Phi}            
\def\n{\nu}             \def\j{\psi}    
\def\r{\varrho}         \def\s{\sigma}  
\def\t{\tau}                  
                     
\def\w{\omega}                            

 \def\LL{{\cal L}}

\def\cl{\centerline}    
\def\ni{\noindent}      \def\pa{\partial}       \def\dd{{\rm d}}

\def\fn#1{\ifcase\noteno\def\fnchr{*}\or\def\fnchr{\dagger}\or\def
	\fnchr{\ddagger}\or\def\fnchr{\rm\S}\or\def\fnchr{\|}\or\def
	\fnchr{\rm\P}\fi\footnote{$^{\fnchr}$} 
	{\scrunch#1\toe}\ifnum\noteno>5\global\advance\noteno by-6\fi
	\global\advance\noteno by 1}
 	\def\scrunch{\baselineskip=11 pt \medrm}
 	\def\toe{\vphantom{$p_\big($}}
	\newcount\noteno
 	% 	 footnote with alternating symbol	%

\def\ffract#1#2{{\textstyle{#1\over#2}}}
\def\fract#1#2{\raise .35 em\hbox{$\scriptstyle#1$}\kern-.25em/
	\kern-.2em\lower .22 em \hbox{$\scriptstyle#2$}}

\def\half{\ffract12} \def\quart{\ffract14}

\def\part#1#2{{\partial#1\over\partial#2}} 
 \def\ref#1{${\vphantom{)}}^#1$}

\def\bbf#1{\setbox0=\hbox{$#1$} \kern-.025em\copy0\kern-\wd0
        \kern.05em\copy0\kern-\wd0 \kern-.025em\raise.0433em\box0}              
                % boldface in math mode.

\def\ref#1{${\,}^{\hbox{\smallrm #1}}$}

\def\Gbar{\raise.13em\hbox{--}\kern-.35em G}
\def\lap{\setbox0=\hbox{$<$}\,\raise .25em\copy0\kern-\wd0\lower.25em\hbox{$\sim$}\,}
\def\glt{\setbox0=\hbox{$>$}\,\raise .25em\copy0\kern-\wd0\lower.25em\hbox{$<$}\,}
\def\gap{\setbox0=\hbox{$>$}\,\raise .25em\copy0\kern-\wd0\lower.25em\hbox{$\sim$}\,}
\def\nc{\medskip\noindent}

%{\nopagenumbers %
{\ }\vglue 1truecm
\rightline{THU-98/33}
\rightline{hep-th/9808154}

\hugeskip
\cl{\bigbf WHEN WAS ASYMPTOTIC FREEDOM DISCOVERED?}\medskip
\cl{\bf or}\medskip
\cl{\bigbf THE REHABILITATION OF QUANTUM FIELD THEORY\fn{Talk 
delivered at the QCD Euroconference 98 on 
Quantum Chromodynamics, Montpellier, July 1998.}}

\hugeskip

\cl{Gerard 't Hooft } \bigskip \cl{Institute for Theoretical Physics}
\cl{University of Utrecht, Princetonplein 5} \cl{3584 CC Utrecht, the
Netherlands} \smallskip\cl{e-mail:  \tt g.thooft@phys.uu.nl} \hugeskip

\ni{\bf Abstract}\Narrower We glance back at the short period of the great
discoveries between 1970 and 1974 that led to the restablishment of Quantum
Field Theory and the discovery of the Standard Model of Elementary
Particles, in particular Quantum Chromodynamics, and ask ourselves 
where we stand now.  \Endnarrower \hugeskip

{\ni\bf 1.  An apology} \nc At meetings such as this one, one should be
looking forward, not back.  But what happened in the crucial period
1970--1974 has been decisive for our field of research, and to set the
record straight is not only of importance for us personally, but also
allows us to learn very important lessons for the future.  Quantum
Chromodynamics as a theory for the strong interactions, and in particular
asymptotic freedom, were not single discoveries made at one single instant.
These insights have a complicated history, and a large number of physicists
contributed here, everyone in his or her own way.\secbreak

{\ni\bf 2.  Prehistory (1947-1970)} \nc In the early '60s, most physicists
did not believe that strong interactions were described by a conventional
quantum field theory.  Much effort went into the attempts to axiomise the
doctrine of relativistic quantum particles strongly interacting among
themselves.  All we had, for sure, was that there should exist an
$S$-matrix, and it obeys all sorts of properties:  unitarity, causality,
crossing symmetry.\ref{1, 2}

Perturbatively renormalizable quantum field theories did exist, and there
were two prototypes:  {\it Quantum Electrodynamics\/} (QED), which had been
very successful in explaining features such as the Lamb shift and the
anomalous magnetic moment of the electron, and $\l\f^4$ theory, describing
a self-interacting scalar field, but which had no obvious application
anywhere in elementary particle physics.  In spite of their remarkable
successes, these theories were considered to be suspect, for various
reasons\ref3.  On hindsight, one could say that this was based mostly on
misundestandings concerning the renormalization group.

In the (former) Soviet Union, Quantum Field Theory was
rejected on the basis of renormalization group arguments.  In particular,
Landau and Pomeranchuk claimed that inadmissible poles would appear in the
ultraviolet region.  This argument appeared to be based on the
K\"allen-Lehmann representation\ref4 of the propagator,
$$D(k^2)=\int_0^\infty{\r(m^2)\dd m^2\over k^2+m^2-i\e}\
;\qquad\r(m^2)\ge0\ , \eqno(2.1)$$ which seemed to imply that the function
now called $\b$-function, had to be positive.

In the West, the Renormalization group had been introduced by A.~Peterman and
E.~Stueckelberg\ref5  in 1953, and it was used by M.~Gell-Mann and F.~Low\ref6 in 1954 to argue
that a renormalizable theory could have a {\it fixed point\/} in the
ultraviolet.  This would be much better than a pole; indeed, the
investigation had been inspired by the (false) expectation that this might
allow one to actually calculate constants of Nature such as the
finestructure constant.  But, here, renormalizable field theories were also
considered to be ugly; they gave the impression that the difficulties, the
infinities, were being ``swept under the rug''.\ref3

Thus, it was widely thought that the real world is not at all described by
a renormalizable quantum field theory (RQFT), and the search for
alternative approaches began.

Experimental evidence showed that strong interactions have a
well-defined symmetry structure.  There is a practically exact $U(2)$
symmetry (isospin and baryon number conservation), a softly broken $SU(3)$
(Gell-Mann's ``Eightfold way"\ref{7, 8}), and a more delicate chiral
$SU(2)\times SU(2)$ that is both spontaneously broken (so that the pions
became massless Goldstone bosons) and explicitly broken (the pions still
have a non-vanishing mass).  It was suspected that the Noether currents
associated with these symmetries were playing an essential role in the
dynamics.  It was attempted to write down rigorous algebras for these
currents, such that their representations would generate the sought-for
$S$-matrix\ref9.  They were combined with unitarity, causality and crossing
symmetry as these were suggested by quantum field theory, but one did not
want to use quantum field theory itself.

A promising new idea was ``duality" (in terms of the Mandelstam variables
$s$, $t$ and $u$ for elastic scattering events)\ref{10}.  A great discovery by
G.  Veneziano\ref{11} was that one could write down simple mathematical
expressions for amplitudes that obeyed this duality requirement.  Since
this duality is somewhat different from what one has in standard quantum
field theory, it was clear why there was much enthusiasm for it.  After
Z.~Koba and H.-B.  Nielsen found how to generalise Veneziano's expressions
to $n$-particle amplitudes,\ref{12}  the physical interpretation was also
discovered, by Nielsen and D.~Fairlie\ref{13}, T.~Goto, L.~Susskind and Y.~Nambu:\ref{14}
the amplitudes are the scattering amplitudes for strings, colliding, merging
and splitting one against the other.

\secbreak {\ni\bf 3.  New models} \nc Yet interest in RQFT did not die
completely.  A monumental beacon was the paper by C.N.~Yang and R.~mills\ref{15} in 1954,
in which they showed that a fundamental principle could be used to
construct impressive field theories:  {\it local Non-Abelian gauge
symmetry}.  Being a direct generalization of QED, this theory appeared to
be renormalizable.  The Yang-Mills Lagrangian was 
$$\LL_{\rm YM}=-\quart G_{\m\n}^a G_{\m\n}^a-\bar\j(\g_\m D_\m +m)\j\,, \eqno(3.1)$$ 
The paper suggested to use this theory to turn isospin symmetry into a local
symmetry, but this clearly could not work, because it would imply the
existence of massless, charged mesons with spin one, a manifest absurdity.

Nevertheless, it was also clear to many researchers that this was a very
deep idea, and, somehow, it should be tried to incorporate it in new
theories for the fundamental interactions. R.~Feynman\ref{16}  was inspired by
it in his first attempts to quantise the gravitational force. S.L.~Glashow\ref{17}, in
1961, used the Yang-Mills Lagrangian as a starting point for an electroweak
theory with $SU(2)\times U(1)$ symmetry, but he had been forced to alter
the theory somewhat in order to accommodate for the mass of the
intermediate vector boson. S.~Weinberg\ref{18} and A.~Salam\ref{19} invoked the Higgs mechanism
to produce a mass term, but they hit upon two problems:  the
renormalization was not understood, and the hadronic sector did not seem to
fit well. M.~Veltman\ref{20}, during the years 1963--1970, used Glashow's approach
and tried to enforce renormalization schemes on it.  He had convinced
himself that there had to be a Yang-Mills structure in the weak
interactions, but abhorred formal theories including the Higgs-Kibble
mechanism.

In the USSR, L.~Faddeev and V.~Popov\ref{21}, E.~Fradkin, and I.V.~Tyutin\ref{22} continued
investigations on pure Yang-Mills and gravity theories.  They generally
avoided renormalization issues, but the renormalization program, for theories
simpler than the Yang-Mills system, and strictly
in perturbative terms, was also continued, by N.N.~Bogolyubov, O.S.~Parasiuk, K.~Hepp,
H.~Lehmann, K.~Symanzik and W.~Zimmermann, among others.

Then, in the '60s, new interest in renormalization arose with the advent of
the $\s$-model.  This was a remarkable new renormalizable model describing
protons, neutrons, pions, and one newly proposed particle, the $\s$.  It
had been invented by Gell-Mann and M.~L\'evy\ref{23} in order to have a realization of
the observed spontaneously broken chiral $SU(2)\times SU(2)$ symmetry.  Its
formal structure was investigated by Symanzik\ref{24}, B.W.~Lee\ref{25} and
J.-L.~Gervais\ref{26}.  but, as its coupling constant, the pion-nucleon interaction
constant $g$, is quite large, one had to attempt to replace the usual
perturbative expansion by something more convergent.  It was
attempted to use Pad\'e approximation techniques to achieve more meaningful
expressions\ref{27}.

\secbreak{\ni\bf 4.  Renormalization}\nc In 1970, we suspected that
explicit mass terms are not allowed in Yang-Mills theories, but we
 still experienced much resistance against the idea of a Higgs mechanism.  
Not only did it seem to be `ugly', in comparison with the `clean' Yang-Mills
concept, the rejection of the Higgs mechanism was also rationalised by
expressing the fear that this would induce far too large contributions to
the gravitational cosmological constant\ref{28}.  Thus, since my then thesis
advisor rejected the Higgs mechanism, and I rejected the introduction of
explicit mass terms, a compromise was agreed upon, and I began my research by investigating the
renormalizability of pure, massless Yang-Mills theories.  It was
established\ref{29}, that renormalizability requires the validity of certain
generalized Ward identities\fn{These identities are now called
Slavnov\ref{30}-Taylor\ref{31} identities, after the authors who rephrased these identities
to have them also hold off-mass shell.  These off-shell versions are
straightforward generalizations, and not directly needed for the
renormalization program.}, which will guarantee unitarity, causality and
renormalizability to be maintained at all orders in perturbation theory.

Once this was realised, it became evident how to do the `massive' case\ref{32}:
one {\it must\/} have a Higgs mechanism if a mass term is desired.
Gauge-invariance then still guarantees unitarity and causality.  Thus, the
Higgs mechanism is far from formal, it is a necessary ingredient for the
electroweak interaction theory.

The new renormalizable electroweak theory enjoyed instant successes.  It
could quickly be extended to encompass the hadrons, because now the {\it
exact rules\/} were understood. The ingredient needed
for this had already been proposed in 1970:  the GIM mechanism\ref{33}.  This
mechanism did require the existence of a new quark flavor, charm, and
furthermore it predicted the existence of neutral components in the weak
interactions.  These `neutral current events' were confirmed in 1973.
Finally, there was what appeared to be a technical detail:  to ensure the
cancellation of dangerous anomalous contributions that would otherwise have
destroyed renormalizability, it was predicited that the number of quark
flavors and lepton species had to be the same.  This was confirmed when the discovery of a
new family of leptons, $\t$ and $\n_\t$, was followed by
that of two new quark flavors:  bottom $b$ and top $t$.

\secbreak {\ni\bf 5.  Confusions about Scaling} \nc So, what happened to
the earlier objections against renormalizable field theories?  It now seems
that these were primarily besed upon confusions concerning the scaling
behavior of quantum field theories.  On hindsight, there had been several
instances in the early days that could have unveiled the misunderstandings.
In 1964, in the USSR, V.S.~Vanyashin and M.V.~Terentyev\ref{34}  found a negative
sign in the charge renormalization of charged vector mesons, but they
attributed this ``absurd'' result to the non-renormalizability of the
theory.  In 1969, Khriplovich\ref{35} correctly 
calculated the charge renormalization of Yang-Mills theories in the Coulomb gauge,
where there are no ghosts. He found the unusual sign, but a connection
with asymptotic freedom was not made, and his important result was not noticed.

\def\forml#1{\setbox1=\hbox{\raise-.55cm\hbox{#1}}\box1}

My own interest in scaling began in 1971.  At first sight, it looked as if
the thing needed for the calculation of the theory's scaling behavior was
nothing but the two-point funcions.  Indeed, they add up to give an
amplitude that appears to be gauge-invariant:  
$$\forml{\epsffile{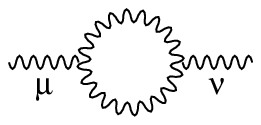}}-\forml{\epsffile{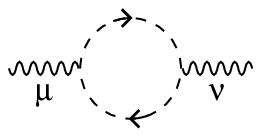}}=\ C\,(k_\m
k_\n-k^2\d_{\m\n})\,, \eqno(5.1)$$ 
This, however, is deceptive.  One might think that gauge invariance now may
be used to deduce an expression of the form $G^a_{\m\n}G^a_{\m\n}$, so that
the renormalization of the coupling constant can be found directly, but
this is false.  Computing diagrams of the form
$$ \forml{\epsffile{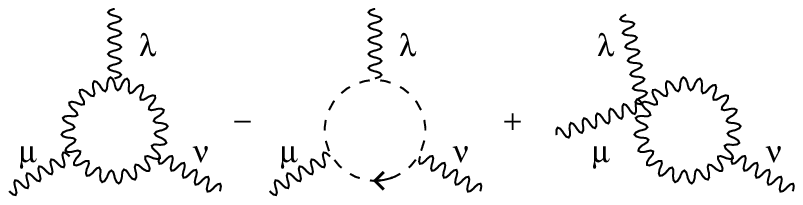}}\matrix{+\ \hbox{permu-}\cr
\hbox{tations,}\cr {\ }\cr} \eqno(5.2)$$ 
shows that their contribution carries an unrelated coefficient.  The reason
for this is that the Feynman rules used to compute these diagrams were
obtained {\it after\/} fixing a gauge condition, such as the Lorenz gauge, so that
then the use of gauge invariance requires more skill.  Thus, to find out
how the coupling constant scales, one must compute field renormalization
effects and the vertex renormalization effects separately.  In the
beginning, these issues were not quite clear, so the exact calculation of
the scaling of the coupling strengths was difficult.  However, the {\it
sign\/} of the effect was never in doubt.  I knew that the coupling had to
decrease at high energies already in 1971, so, in any case, I was immune
to the Landau objections.

Meanwhile, experiments were moving towards the higher energy domain, and
here a peculiar scaling phenomenon was observed. The general nature of
this behaviour was investigated by J.D.  Bjorken\ref{36}.  This
experimental discovery became to be known as ``Bjorken scaling''.  It was
subsequently interpreted by Feynman\ref{37} as if, at short distances,
constituent particles inside hadrons move comparatively freely.  He called
these constituents `partons'.  It would later turn out to be a good insight
{\it not\/} to call them `quarks', because now we know that partons can be
both quarks and gluons.

In the late '60s, C.G.~Callan\ref{38} and K.~Symanzik\ref{39} independently
found equations obeyed by the renormalized Green functions that follow from
their scaling behaviour, now known as the Callan-Symanzik equations.  They
explicitly constructed their equations only for QED and $\l\f^4$ theory,
but they expected them to have universal validity.  In particular, the $\b$
functions were found to be positively valued, and this too was thought to
be a universal feature, valid for {\it all\/} quantum field theories,
Abelian as well as non-Abelian ones.  For this reason, it was generally felt
that {\it no quantum field theory will ever explain Bjorken scaling}\ref{40}.
I could not understand why people were saying this because I knew about the
beautiful scaling behaviour of non-Abelian gauge theories.  Suspecting that
this feature should be known by now by the experts on the subject of
scaling, I did not speak up louder.  When I mentioned my suspicions to
Veltman, that a pure $SU(3)$ gauge theory could be all what is needed to
describe the forces between quarks, he warned me that no-one would take
such an idea seriously as long as it could not be explained why quarks
cannot be isolated one from another.

By 1972, I had calculated the scaling behavior, and I wrote it in the form
$${\m\dd\over\dd\m}
g^2={g^4\over8\pi^2}\big(-\ffract{11}3\,C_1+\ffract23\,C_2
+\ffract16\,C_3\big)\,,\eqno(5.3)$$ in a theory in which the gauge group
has a Casimir operator $C_1$ and where fermions have a Casimir operator
$C_2$.  Scalars contribute with the Casimir operator $C_3$.  In the $SU(2)$
case this would mean that with less than 11 fermions in the elementary
representation (and no scalars) the $\b$ function would be negative, and
for $SU(3)$ the critical number would be $33/2$.

In June, 1972, a small meeting was organised by C.P.~Korthals Altes in Marseille.  
Symanzik  explained that he worked on $\l\phi^4$ theory with
negative $\l$, because of its interesting scaling properties\ref{41}.  It
was unfortunate, but, in Symanzik's opinion, perhaps not insuperable, 
that the only models with
such properties that he could think of were theories with a negative
squared coupling constant, $g^2<0$, or $\l<0$. He explained that perhaps this problem
could be cured by non-perturbative effects, which at that time were very
badly understood.  I announced at that meeting my finding that the
coefficient determining the running of the coupling strength, that he
called $\b(g^2)$, for non-Abelian gauge theories is negative, and I wrote down
Eq.~(5.3) on the blackboard.  Symanzik was
surprised and skeptical.  ``If this is true, it will be very important, and
you should publish this result quickly, and if you won't, somebody else
will,'' he said. I did not follow his advice. A long calculation on
quantum gravity with Veltman had to be finished first.

When Veltman and I had worked out many examples of gauge models\ref{42}, we knew
how to compute the renormalization counter terms:  
$$\D \LL_{\rm Dim.Regul.}={1\over8\pi^2(4-n)}\cdot\quart G_{\m\n}^a
G_{\m\n}^a\big(-\ffract{11}3\,C_1+\ffract23\,C_2+\ffract16\,C_3\big)\,.
\eqno(5.4)$$ 
Clearly, this had to be related to the scaling behavior found
in Eq.~(5.3).  The basis for this relation was quickly found\ref{43}
Furthermore, it was found how to use background field methods for a quick
derivation of Eq.~(5.4), and hence also (5.3).  With these methods, we
could limit ourselves to diagrams of the kind used in Eq.~(5.1), and ignore
the ones of (5.2), by making use of slightly altered Feynman rules\ref{44}.

 As for
what happened in the States, and how the sign of the $\b$ function for
non-Abelian gauge theories was discovered, I leave to D.~Gross to account.

When I heard about the publications by H.D.~Politzer\ref{45}, D.~Gross and
F.~Wilczek\ref{46}, I was surprised, not about the result, which I had known all
along, but about the stir they caused.  And, finally, people began to talk
about a pure $SU(3)$ theory for quarks.  Well, I thought, they still have
nothing; as Veltman had taught me, what really matters is the quark
confinement problem.  No real advances had been made yet.  I set out to try
to understand confinement from first principles.  QCD had only come half-way.

In 1973, Gross and Coleman showed furthermore that asymptotic freedom would {\it
never\/} occur in theories that only contain fermions and scalars.\ref{47} 
First, this result could be generalized by writing down the
complete algebra for all one-loop beta functions\fn{Gross and Coleman had
included two-loop diagrams for the scalar field renormalization, but these
are not needed for the result.}.

In all renormalizable theories, one can write the basic Lagrangian as
$$\eqalign{\LL^{\rm inv}&= -\quart\,G_{\m\n}G_{\m\n}-\half(D\f)^2-V(\f)\cr
&-\bar\j\,\g D\j-\bar\j\big(S(\f)+i\g_5 P(\f)\big)\j\,,}\eqno(5.5)$$ where
$V(\f)$ is at most a quartic polynomial in the scalar field(s) $\f_i$, and
$S$ and $P$ are linear expressions in $\f_i$, which may be anything as long
as the total Lagrangian is kept gauge invariant. In (5.5), the fields $\f_i$
are written as real fields, $\j_i$ and $\overline{\j_i}$ are complex.

\def\Tr{{\rm Tr}} The one-loop counter Lagrangian is then\ref{44}:  
$$\D\LL={1\over
8\pi^2(4-n)}\Big\{-\quart\,G_{\m\n}G_{\m\n}\big(\ffract{11}3\,C_1
-\ffract23\,C_2-\ffract16\,C_3\big)-\D V-\bar\j(\D S+i\g_5\D
P)\j\Big\}\,,\eqno(5.6)$$ with, writing $S+iP\equiv W$, $$\eqalign{\D V&= \quart(\pa_i\pa_j
V)^2+\ffract32\pa_i V(T^2\f)_i+\ffract34 (\f T^a T^b\f)^2\cr &+\f_i
V_j\,\Tr\big(S_i S_j+P_i P_j\big)-\Tr\big(S^2+P^2\big)^2 +\Tr[S,P]^2\,;\cr
 \D W&=\quart W_iW^*_i W+\quart WW_i^* W_i+W_i W^*W_i\cr
&+\ffract32(U_R^2 W)+\ffract32 W(U_L)^2+W_i\f_j\Tr\big(S_iS_j+P_iP_j\big)\
,} \eqno(5.7)$$ where $T^a$, $U^a_R$ and $U^a_L$ are the generators of
gauge transformations on $\F_i$, $\j_{i,R}$ and $\j_{i,L}$, which determine
the coupling strengths of these fields to the gauge field. $V_i=\pa_i V$
stands for differentiation of $V(\f)$ with respect to the fields $\f_i$.

Carefully studying the signs of these expressions, one finds in general:
$$\eqalignno{\b(g)&=(-R_1+R_2N_f+R_3N_s)g^3\ ;&(5.8)\cr
\b(Y)&=R_4Y^3-R_5g^2Y\ ;&(5.9)\cr \b(\l)&=R_6\l^2-R_7
g^2\l+R_8g^4+R_9Y^2\l-R_{10}Y^4\ ,&(5.10)}$$ where $g$ stands short for all
gauge field coupling constants, $Y$ for all Yukawa couplings, and $\l$ for
all quartic scalar field self-couplings.  $R_{1-10}$ are all positive
constants.  The relative strengths of these constants depends on further
details such as the nature of the gauge group representations and the
multiplicities of these fields, and can vary quite a bit.  In general, one
can find several choices of gauge field theories and coupling constants such that all
constants run to zero at infinite energies.  

\secbreak {\ni\bf 6.  Color $SU(3)$ -- the successes of QCD.}  
\nc The early quark theory suffered from an apparent contradiction
with the spin-statistics theorem, since the baryonic states appeared
to be described by totally symmetric quark wave functions, instead of
the expected antisymmetric ones. In 1964, Greenberg\ref{48} suggested a
solution in terms of a new kind of statistics called `parastatistics'.
The first ideas of an internal
`color' gauge group causing the confining forces between quarks go back to
Han and Nambu\ref{49} in 1965, who produced models in
which electric charge was linked to color in such a way that physical
charges would be integral, in what would nowadays be called a Higgs theory.
The idea is not as far away from the truth as one might think.  If there
existed scalar quarks, these could automatically be cast in a form that
would make QCD look like a Higgs theory, without major departures from its
present appearance.  At the time that Gross and Wilczek wrote their
papers\ref{50}, there were also proposals by H.~Fritzsch, Gell-Mann and
H.~Leutwyler\ref{51}, who investigated pure $SU(3)$ color theories for quarks,
and they recognised a serious problem (the $U(1)$ problem, to be discussed
in Sect.~7).  There were also similar suggestions by H.~Lipkin\ref{52}.

None of these papers had a satisfactory answer to the question why quarks
are confined.  But then several key developments took place.  The {\it dual
resonance models\/}\ref{10, 11} gave a first clue:  mesons appear to behave as little
stretches of strings, with quarks at their end points.  This most
beautifully squared with the observed linear relation between the
mass-squared and the angular momentum of the heavy mesonic resonances, the
so-called Regge trajectories\fn{The work of J.~Chew\ref{1},
S.~Frautschi\ref{53} and V.~Gribov\ref{54} should be mentioned here.}.  It
suggested that the colored forces between quarks form {\it vortex lines\/}
that connect color charges.  How can one understand why these are formed?
This is a long-distance problem, and precisely at these long distances (of
a fermi and more), the coupling strength grows so large that perturbative
arguments will be meaningless.

Independently, Nielsen and P.~Olesen\ref{55} in Copenhagen, and
B.~Zumino\ref{56} at CERN, discovered a new and extremely important feature
in an {\it Abelian\/} Higgs theory:  this model would allow for the
existence of a magnetic vortex line that would have several features in
common with the dual string:  it is unbreakable, its mass-energy content is
entirely due to the tension force, and it exhibits an internal Lorentz
invariance (for boosts in the direction of the string).  If quarks would be
magnetic monopoles with respect to this spontaneously broken $U(1)$ force,
you would get exactly the model you want.  In fact, this model is nearly
identical to much older models for superconducting material in which the
Meissner effect is active.  The vortices are just the well-known Abrikosov
vortices formed by magnetic fields that succeed in penetrating into a
superconductor.

But quarks are not monopoles.  This theory was not QCD, and so it would not
explain Bjorken scaling and the symmetric $SU(3)$ representations of the
hadrons.  What happens if one replaces the Abelian $U(1)$ theory by a
non-Abelian theory with Higgs mechanism?  Does that also allow for
Abrikosov vortices?  I asked myself this question, and ran into a puzzle:
there would be vortices, but they would differ from the Abelian ones in
that now they can break.  The resolution of the puzzle was revolutionary:
the vortices break because they can create pairs of magnetic monopoles and
antimonopoles.  It was not known yet that magnetic monopoles can exist in
fairly ordinary non-Abelian gauge models!\ref{57}.  Independently, but a
couple of months later, this discovery was made in Moscow:  A.~Polyakov had
constructed the same gauge field solution\ref{58}, and it was remarked by
L.  Ok\'un, that the object carries a single magnetic charge.

However, it looked as if the monopole was a side-track.  By itself it did
not bring us closer to understanding quark confinement. Yet it would later
turn out to be a crucial ingredient.

First, something else happened.  K.~Wilson was studying the renormalization
group for non-Abelian gauge theories restricted on a space-time lattice.
His interest was primarily in the domain of computer simulations.  But he
made a crucial observation\ref{59}:  if, for practical, technical reasons,
you restrict (Euclidean) space
and time to form a lattice, you can perform a new perturbative expansion:
the $1/g^2$ expansion.  This large coupling constant expansion becomes
accurate at large distance scales.  Surprisingly, one can read off right
away that the only states that one can then talk about are gauge-invariant
(that is, colorless) states, formed by quarks that had to be connected one to
another by line segments of the lattice, exactly as in the dual string
theory!  In other words, quark confinement is evident {\it and absolute\/}
in the large coupling constant expansion!

In my recollection, this was the first indication that quark confinement is
absolute, quarks can never escape from each other.  It looks as if, for a
theory on a lattice, when the coupling constant is varied from small to
large, a {\it phase transition\/} occurs.  QCD is said to be in the {\it
confinement phase\/}.  Yet, in the continuum theory, the interactions are
generated at very tiny distance scales, where the coupling constant is
weak, and then, by cumulative effects, they generate larger couplings at
large distances.  Does this justify a large coupling constant expansion for
continuum theories?  Not really.  We were coming closer, but the exact
nature of the confinement mechanism had not yet been explained
convincingly.

Because of Dirac's relation\ref{60} between the fundamental electric charge
quantum $e$ and the magnetic charge quantum $g_m$, $$g_m\,e=2\pi n\hbar
c\,,$$ where $n$ is an integer, the interchange electric--magnetic also
corresponds to a small coupling--large coupling interchange. 
Therefore, to understand confinement, it should be a good
idea to interchange the electrically charged particles with the magnetic
monopoles in the theory.  If, subsequently, a Higgs mechanism is assumed to
take place among the magnetic monopoles, we obtain the dual analogue of the
older model of Nielsen, Olesen and Zumino.  The magnetic vortex in that
model would now manifest itself as an electric vortex line, and the role of
the monopoles in that model would now be played by the quarks.  They will
be held together absolutely and permanently by these vortices.  Bingo, this
is permanent quark confinement.\ref{61, 62}

The next step would be to reformulate this mechanism in more precise terms.
The description of phase transitions using dual transformations turned out
to be related to what is done in statistical systems.\ref{63}
But also, one had to rephrase carefully the gauge constraints of the
non-Abelian theory.  It had been noted by Gribov\ref{64}, that the usual
procedure for fixing the gauge degree of freedom in an Abelian gauge theory
leads to a complication when applied to the non-Abelian case:  the Lorenz
condition does {\it not\/} fix the gauge freedom completely; there is an
ambiguity left.  This ambiguity is harmless as long as one sticks to a
perturbative expansion.  There, we know exactly what kind of non-local
effects are caused by gauge fixing:  it is the Faddeev-Popov ghosts.  

But in attempts to go beyond perturbation expansion, Gribov's ambiguity is
important.  It is then more difficult to distinguish physical states
from ghosts.  Ghosts arise as soon as the gauge fixing procedure requires
the solution of differental equations; their kernels are the ghost
propagators.  If one wants to avoid ghosts altogether, one must use a {\it
local, non-propagating\/} gauge fixing procedure.  Indeed, it is possible
exactly to give a gauge constraint that removes {\it the non-Abelian
parts\/} of the local gauge invariance, without any ambiguity\ref{65}.
After such a gauge fixing procedure, one is left with an apparently Abelian
effective theory, {\it with\/} in addition to the electrically charged
particles, {\it magnetic monopoles\/}.  They emerge as singularities due to
the gauge fixing.  Since the Abelian gauge group that survives is the
Cartan subgroup of the original gauge group (the largest Abelian subgroup),
we call this procedure the `maximal Abelian gauge fixing procedure', or
`the Abelian projection'.

In the maximal Abelian gauge, we find `color-electrically' charged
particles, `color-magnetically' charged particles, each with finite masses
depending on the dynamical features of the system, and massless $U(1)$
`color-photons' (gluons), that couple to the `color-electric' and
`color-magnetic' particles in a totally symmetric fashion.

All one is now left with to do, is to invoke the Higgs mechanism in this
color regime.  If the Higgs mechanism applies to the color-electrically
charged particles, then these will be free, with short-range interactions
only, since they exchange gluon-photons that now have become massive.  The
color-magnetic particles will be held together eternally by color-magnetic
vortex lines.  But the effective model in the maximal Abelian gauge equally
well allows us to assume the Higgs mechanism to work on the
color-magnetically charged particles, in which case the color-electrically
charged particles are held together by color-electric vortices, being the
dual strings.  Whether the color-electric, or the color-magnetic Higgs
mechanism takes place, or perhaps neither, depends on the relative strength
of the color-electric and the color-magnetic charges.  Due to asymptotic
freedom, one generally expects the color-electric charges to become large
when it is attempted to separate them by large distances; the
color-magnetic charges will become small, with a tendency to be screened
completely.  It is natural to assume that the latter undergo Bose
condensation, just as in super conductors, and then one has permanent quark
confinement.

A daring experiment was first carried out by M.~Creutz {\it et al\/}, on a
computer\ref{66}:  they took a lattice in four-dimensional Euclidean space
and defined the QCD field variables on this lattice.  Because of the
complexity of the problem, and the limited capacity of these early
computers, the lattices had to be very modest in size.  Nevertheless, their
computer simulations showed clear evidence of a transition towards a
confining phase.  Nowadays, by using much larger and faster computers, as
well as much improved analytical procedures, this transition can be studied
in much more detail.  One has also been able to impose maximal Abelian
gauge conditions on the leading color field configurations that were
obtained, and quite generally, the picture of Bose-condensing magnetic
monopoles has been confirmed\ref{67}.  Today, it is fair to say that quark
confinement is no longer seen as a deep puzzle in quantum field theory.  We
therefore understand why protons did not break into quarks in the ISR
experiments at CERN\ref{68}. If a color vortex ever breaks, it is because of
quark-antiquark pair formation, so that hadrons are always kept color
neutral.  A vortex breaking without quark pair formation would require
large-scale structures ({\it i.e.}, low energy excitations) of a kind that
directly contradict the lattice simulations.  Therefore, we do not expect
that experimentalists will ever detect free quarks.

Experimental support for QCD then came in numerous ways.  Notably, the
quark and gluon substructures became clearly recognisable as `jets', and
the slow variations of the running coupling constants were confirmed.

\secbreak {\ni\bf 7.  The $U(1)$ problem} \nc

A serious problem, however, was cause for concern.  This was the so-called
$U(1)$ or $\eta$-$\eta'$ problem.  The {\it explicit\/} breakdown of chiral
$U(1)$ symmetries had been noted as an `anomaly' in quantum field
theories\ref{69}:  the $\pi^0$ decay into two photons, for instance, would
have been forbidden because of a chiral conservation law, but the process
nevertheless takes place.  It also is found to take place if one performs a
simple-minded calculation, either using an intermediate proton-antiproton
state (Fig.~1a), or quark-anti-quark states (Fig.~1b).

\midinsert\cl{\epsffile{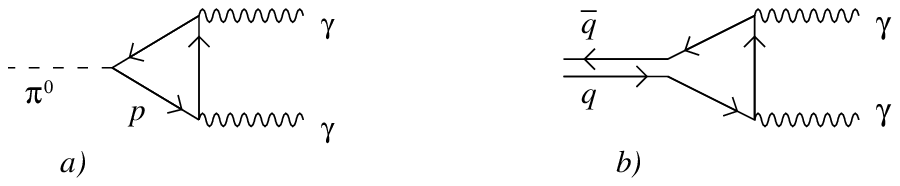}}
\cl{Fig.~1. The $U(1)$ anomaly, a) in terms of virtual protons, b)~in terms of virtual quarks.}
\endinsert

But what happened to this symmetry in QCD was not understood.  The problem
was recognised already by the earliest authors, Fritzsch, Gell-Mann and
Leutwyler\ref{51}.  The QCD Lagrangian correctly reflects all known
symmetries of the strong interactions, and the hadron spectrum is
reasonably well explained by extending these symmetries to an approximate
$U(6)$ symmetry spanned by the $d$, $u$ and $s$ quark each of which may
have their spins up or down.

But the QCD Lagrangian also shows a chiral $U(2)\times U(2)$ symmetry, and
consequently, it suggests the presence of an additional axial $U(1)$
current that is approximately conserved.  This should imply that a
pseudoscalar particle (the quantum numbers are those of the $\eta$
particle) should exist that is about as light as the three pions.  Yet, the
$\eta$ is considerably heavier:  549 MeV, to be compared with the 135 MeV
of a pion.  For similar reasons, one should expect a ninth pseudoscalar
boson that is as light as the kaons.  The only candidate for this would be
one of the heavy mesons, originally called the $X^0$~meson.  When it was
discovered that this meson can decay into two photons\ref{70}, and
therefore had to be a pseudoscalar, it was renamed as $\eta'$.  However,
this $\eta'$ was known to be as heavy as 958 MeV, and therefore it appeared
to be impossible to accommodate for the large chiral $U(1)$ symmetry
breaking by modifying the Lagrangian.  Also, the decay ratios -- especially
the radiative decays -- of $\eta$ and $\eta'$ appeared to be anomalous, in
the sense that they did not obey theorems from chiral algebra.\ref{71}

Clearly, there is an anomaly akin to the Adler-Bell-Jackiw anomaly.
The divergence of the axial vector current is corrected by quantum effects:
$$\pa_\m J^A_\m={g^2\over16\pi^2}\e_{\m\n\a\b}\,{\rm
Tr}\,G_{\m\n}G_{\a\b}\,, \eqno(7.1)$$ where $J^A_\m$ is the axial vector
current, $G_{\m\n}$ the Yang Mills gluon field, and $g$ the strong coupling
constant.  So, the axial current is not conserved.  Then, what is the
problem?

The problem was that, in turn, the r.h.s.  of this anomaly equation can
also be written as a divergence:
$$\e_{\m\n\a\b}\,\Tr\,G_{\m\n}G_{\a\b}=\pa_\m K_\m \,,\eqno(7.2)$$ where
$K_\m$ is the Chern-Simons current.  $K_\m$ is not gauge-invariant, but it
appeared that the latter equation would be sufficient to render the $\eta$
particle as light as the pions.  Why is the $\eta$ so heavy?

There were other, related problems with the $\eta$ and $\eta'$ particles:
their mixing.  Whereas the direct experimental determination of the
$\w$-$\phi$ mixing\ref{72} allowed to conclude that in the octet of
vector mesons, $\w$ and $\f$ mix in accordance to their quark contents:
$$\w=\ffract{1}{\sqrt2}(u\bar u+d\bar d)\ ,\qquad\f=s\bar s\ ,\eqno(7.3)$$
the $\eta$ particle is strongly mixed with the strange quarks, and $\eta'$
is nearly an $SU(3)$ singlet:  $$\eta\approx\ffract{1}{\sqrt3}(u\bar
u+d\bar d-s\bar s)\ ,\qquad \eta'\approx\ffract{1}{\sqrt6}(u\bar u+d\bar
d+2s\bar s)\ .\eqno(7.4)$$ Whence this strong mixing?

Several authors came with possible cures. J.~Kogut and L.~Susskind\ref{73}
suggested that the resolution came from the quark confinement mechanism,
and proposed a subtle procedure involving double poles in the gluon
propagator.  Weinberg\ref{74} also suggested that, somehow, the would-be
Goldstone boson should be considered as a ghost, cancelling other ghosts
with opposite metrics.  My own attitude\ref{61} was that, since $K_\m$ is
not gauge-invariant, it does not obey the boundary conditions required to
allow one to do partial integrations, so that it was illegal to deduce the
presence of a light pseudoscalar.

Just as it was the case for the confinement problem, the resolution to this
$U(1)$ problem was to be found in the very special topological structure of
the non-Abelian forces.  In 1975, a topologically non-trivial field
configuration in four-dimensional Euclidean space was described by four
Russians, A.A.~Belavin, A.M.~Polyakow, A.S.~Schwarz and
Yu.S.~Tyupkin\ref{75}.  It was a localized configuration that featured a
fixed value for the integral $$\int\dd^4x\,\e_{\m\n\a\b}\,\Tr
\,G_{\m\n}G_{\a\b}={32\pi^2\over g^2}\,.  \eqno(7.5)$$ The importance of
this finding is that, since the solution is localized, it obeys all
physically reasonable boundary conditions, and yet, the integral does not
vanish.  Therefore, the Chern-Simons current does not vanish at infinity.
Certainly, this thing had to play a role in the violation of the $U(1)$
symmetry.

This field configuration, localized in space as well as in time, was to be
called ``instanton'' later\ref{76}.  Instantons are sinks and sources for
the chiral current.  This should mean that chirally charged fermions are
created and destroyed by instantons.  How does this mechanism work?

It was discovered that near an instanton, the Dirac equation shows special
solutions, which are fermionic modes with vanishing action\ref{76}.  This
means that the contribution of fermions to the vacuum-to-vacuum amplitude
turns this amplitude to zero.  Only amplitudes in which the instanton
creates or destroys chiral fermions are unequal to zero.  The physical
interpretation of this was elaborated further by Russian investigators, by
R.~Jackiw, C.~Nohl and C.~Rebbi\ref{77}, and by C.~Callan, R.~Dashen and
D.~Gross\ref{78}. Instantons are tunnelling events.  Gauge field
configurations tunnel into other configurations connected to the previous
ones by topologically non-trivial gauge transformations.  During this
tunnelling process, one of the energy levels produced by the Dirac equation
switches the sign of its energy.  Thus, chiral fermions can pop up from the
Dirac sea, or disappear into it.  In a properly renormalized theory, the
number of states in the Dirac sea is precisely defined, and adding or
subtracting one state could imply the creation or destruction of an
antiparticle.  This is why the original Adler-Bell-Jackiw anomaly was first
found to be the result of carefully renormalizing the theory.

With these findings, effective field theories could be written down in such
a way that the contributions from instantons could be taken into account as
extra terms in the Lagrangian.  These terms aptly produce the required mass
terms for the $\eta$ and the $\eta'$, although it should be admitted that
quantitative agreement is difficult to come by; the calculations are
exceptionally complex and involve the cancellation of many large and small
numerical coefficients against each other.  But it is generally agreed upon
(with a few exceptions\ref{79}), that the $\eta$
and $\eta'$ particles behave as they should\ref{80} in QCD.

It was after the resolution of the confinement problem and the $U(1)$ problem,
that QCD could be accepted as a viable theory. Strong experimental support
was the observation of jet events, which clearly reflect the short distance structure,
and the observation of the running coupling constant. 
  
\secbreak {\ni\bf 8.  New developments still needed}\nc QCD is in a very
good state theoretically as well as experimentally.  Yet, there is still
very much to be desired, first of all, a more reliable on-shell
calculational technique.  A very promising result was that the
theory simplifies if the number of colors, $N_c$, whose physical value is
three, is sent to infinity instead.  One can subsequently consider the
corrections in powers of $1/N_c$.  The advantage of such an expansion would
be that, already at $N_c=\infty$ the spectrum of physical states is
suspected to be very similar to the real spectrum; only the mesons in this
limit are non-interacting.  So, we suspect in the $N_c\ra\infty$ limit an
effective meson theory.

Unfortunately, this limiting theory has not yet allowed a rigorous
mathematical description.  We suspect it to be very similar to string
theories, but the desired string theory has not been found.  very promising
recent developments suggest that progress here is possible.\ref{81}  Also, the
author recently reinvestigated some aspects of the ``planar'' diagram
expansion required by this theory.  We suspect quark confinement to be a
fundamental property of planar diagrams, related to divergences in the
planar diagram expansion.\ref{82}
\secbreak\ni{\bf References}\nc\parskip=3pt  
 
\item{1.} G.~Chew, in {\it The Analytic S-Matrix}  (W.A.~Benjamin Inc, 1966).
\item{2.} M.~Goldberger, in {\it The Quantum Theory of Fields: The 12th
Solvay Conference}, 1961 (Interscience, New York).
\item{3.} See the arguments raised in: R.P.~Feynman, in {\it The Quantum 
Theory of Fields -- The 12th
Solvay Conference}, 1961 (Interscience, New York). See also:
T.Y.~Cao and S.S.~Schweber, {\it The Conceptual Foundations and
Philosophical Aspects of Renormalization Theory}, {\it Synthese  \bf 97} (1993)
33, Kluwer Academic Publishers, The Netherlands.
\item{4.} G.~K\"allen, Helv.~Phys.~Acta {\bf 23} (1950) 201, {\it ibid.}
{\bf 25} (1952) 417; H.~Lehmann,
Nuovo Cimento {\bf 11} (1954) 342.
\item{5.} E.C.G.~Stueckelberg and A.~Peterman, {\it Helv. Phys. Acta  
\bf  26} (1953) 499; N.N.~Bogoliubov and D.V.~Shirkov, {\it 
Introduction  to  the  theory  of quantized fields }
(Interscience, New York, 1959);
A.~Peterman, {\it Phys. Reports  \bf  53c} (1979) 157.
\item{6.} M.~Gell-Mann and F.~Low, {\it Phys. Rev.  \bf  95} (1954) 1300.
\item{7.} M.~Gell-Mann, ``The Eightfold Way - A Theory of Strong-Interaction
Symmetry'', Calif. Inst. Technol. Synchrotron Lab. Report {\bf 20} (1961);
Y.~Ne'eman, {\it Nucl. Phys. \bf 26} (1961) 222; see also M.~Gell-Mann and
Y.~Ne'eman, ``The Eightfold Way'', Benjamin, New York (1964).
\item{8.} M.~Gell-Mann, {\it Phys. Lett. \bf 8} (1964) 214; 
G.~Zweig, Erice Lecture 1964, in {\it Symmetries in Elementary Particle Physics}, 
A.~Zichichi, Ed., Academic Press, New York, London, 1965, and 
CERN Report No. TH 401, 4R12 (1964), unpublished.
\item{9.} M.~Gell-Mann, {\it Physics \bf 1} (1964) 63;  
 R.~Dashen and M.~Gell-Mann, {\it Phys. Rev. Lett. \bf 17} (1966) 340;
S.~Adler, {\it Phys. Rev. \bf 140 B} (1965) 736;
W.I.~Weisberger, {\it Phys. Rev. \bf 143 B} (1966) 1302.
\item{10.} V.~Alessandrini, {\it et al, Physics Reports \bf 1C} (1971) 269;
J.H.~Schwarz, {\it Physics Reports \bf 8C} (1973) 270;
 S.~Mandelstam, {\it Physics Reports.  \bf 13C} (1974) 259.
\item{11.} G.~Veneziano, {\it Nuovo Cim \bf 57A} (1968) 190.
\item{12.} Z.~Koba and H.B.~Nielsen, {\it Nucl. Phys.  \bf B10} (1969) 633,
{\it ibid.  \bf B12} (1969) 517; {\bf B17} (1970) 206; {\it
Z.~Phys.  \bf 229} (1969) 243.
\item{13.} H.B.~Nielsen, "An Almost Physical Interpretation of the Integrand of the 
  n-point Veneziano model, {\it XV Int.  Conf.  on  High  Energy  Physics}, 
  Kiev, USSR, 1970; Nordita Report (1969), unpublished;
 D.B.~Fairlie and H.B.~Nielsen, {\it Nucl. Phys. \bf B20} (1970) 637.
\item{14.} H.B.~Nielsen and L.~Susskind, CERN preprint TH 1230 (1970), Y.~Nambu, 
 Proc. Int. Conf. on {\it Symmetries and Quark models},
Wayne state Univ. (1969); Lectures at the Copenhagen Summer Symposium (1970);
T.~Goto, Progr. Theor. Phys. 46 (1971) 1560;
	L.~Susskind, {\it Nuovo Cimento \bf 69A} (1970) 457; {\it Phys. Rev \bf 1} (1970) 1182.
\item{15.} C.N. Yang and R.L. Mills, Phys. Rev. {\bf  96} (1954) 191, {\it
see also:\/} R. Shaw,  Cambridge   Ph.D. Thesis, unpublished. 
\item{16.} R.P.~Feynman, {\it Acta Phys. Polonica \bf 24} (1963) 697.
\item{17.} S.L.~Glashow, Nucl. Phys. {\bf 22} (1961) 579.
\item{18.} S.~Weinberg, {\it Phys. Rev. Lett. \bf 19} (1967) 1264.
\item{19.} A.~Salam and J.C.~Ward, Phys. Lett. {\bf 13} (1964) 168,
	A.~Salam, Nobel Symposium 1968, ed. N.~Svartholm.
\item{20.} M.~Veltman, {\it Physica \bf 29} (1963) 186,
{\it  Nucl. Phys. \bf B7} (1968) 637; J.~Reiff and  M.~Veltman, {\it Nucl. 
     Phys. \bf B13} (1969) 545; M.~Veltman, {\it Nucl. Phys. \bf B21} (1970) 288;
H.~van Dam and M.~Veltman, {\it Nucl. Phys. \bf B22} (1970) 397.
\item{21.} L.D.~Faddeev and V.N.~Popov, {\it Phys. Lett. \bf 25B} (1967) 29;
L.D.~Faddeev, {\it Theor. and Math. Phys. \bf 1} (1969) 3 (in Russian),
 {\it Theor. and Math. Phys. \bf 1} (1969) 1 (Engl. transl).
\item{22.} E.S.~Fradkin and I.V.~Tyutin, {\it Phys. Rev. \bf  D2} (1970) 2841.
\item{23.} M.~Gell-Mann and M.~L\'evy, {\it  Nuovo Cim.  \bf  16} (1960) 705.
\item{24.} K.~Symanzik, Carg\`ese lectures, July 1970.
\item{25.} B.W.~Lee, "Chiral Dynamics", Gordon and Breach, New York (1972)
\item{26.} J.-L.~Gervais and B.W.~Lee, {\it Nucl. Phys. \bf B12} (1969) 627;
J.-L.~Gervais, Carg\`ese lectures, July 1970.
\item{27.} Various authors, Carg\`ese Summer Institute, July 1970.
\item{28.} M.~Veltman, personal communication.
\item{29.} G.~'t~Hooft, {\it Nucl.~Phys. \bf B33} (1971) 173.
\item{30.} A.~Slavnov, {\it Theor. Math. Phys. \bf 10} (1972) 153 (in Russian), 
{\it Theor.  Math.  Phys. \bf 10} (1972) 99 (Engl. Transl.)
\item{31.} J.C.~Taylor, {\it Nucl. Phys. \bf B33} (1971) 436.
\item{32.} G.~'t~Hooft, {\it Nucl. Phys.  \bf B35} (1971) 167.
\item{33.} S.L. Glashow, J. Iliopoulos and L. Maiani, Phys. 
Rev. {\bf  D2} (1970) 1285 
\item{34.} V.S.~Vanyashin and M.V.~Terentyev, {\it Zh.Eksp.Teor.Phys.
\bf 48} (1965) 565 [{\it Sov.Phys. JETP \bf 21} (1965)]
\item{35.} I.B. Khriplovich, {\it Yad.Fiz. \bf 10} (1969) 409
[{\it Sov.J.Nucl.Phys. \bf 10} (1969)]
\item{36.}  J.D.~Bjorken, {\it Phys. Rev. \bf 179} (1969) 1547.
\item{37.} R.P.~Feynman, {\it Phys. Rev. Lett. \bf 23} (1969) 337.
\item{38.} C.G.~Callan, {\it Phys. Rev. \bf  D2} (1970) 1541.
\item{39.} K.~Symanzik, {\it Commun. Math. Phys.  \bf 16} (1970) 48;
{\it ibid. \bf 18} (1970) 227, {\it ibid. \bf 23} (1971) 49.
\item{40.} D.J.~Gross, personal communication (1971); see also:
D.J.~Gross, in {\it The Rise of the Standard Model},
Cambridge Univ. Press (1997), p. 199.
\item{41.} K.~Symanzik, in Proc. Marseille Conf. 19-23 June 1972, ed. C.P. Korthals 
     Altes; {\it id.},  {\it Nuovo Cim. Lett.  \bf 6} (1973) 77.   
\item{42.} G.~'t~Hooft and M.~Veltman, {\it Nucl. Phys. \bf B44} (1972) 189;
{\it ibid. \bf B50} (1972) 318, and in: Proceedings  of  the  Colloquium  on 
     Renormalization of Yang-Mills Fields, Marseille, June 19-23, 1972.
\item{43.} G.~'t~Hooft, {\it Nucl. Phys.  \bf B61} (1973) 455. 
\item{44.} G.~'t~Hooft, {\it Nucl. Phys.  \bf B62} (1973) 444. 
\item{45.} H.D.~Politzer, {\it Phys. Rev. Lett. \bf 30} (1973) 1346;
\item{46.} D.J.~Gross and F.~Wilczek, {\it Phys.~Rev.~Lett. \bf 30} (1973) 1343; 
\item{47.} S.~Coleman and D.J.~Gross, 
{\it Phys.~Rev.~Lett. \bf 31} (1973) 851.
\item{48.} O.W.~Greenberg, {\it Phys. Rev. Lett. \bf 13} (1964) 598.
\item{49.} M.Y.~Han and Y.~Nambu, {\it Phys. Rev. \bf 139 B} (1965) 1006.
\item{50.} D.J.~Gross and F.~Wilczek, {\it Phys.~Rev. \bf D8} (1973) 3633;
{\it ibid. \bf D9} (1974) 980.
\item{51.} H.~Fritzsch, M.~Gell-Mann and H.~Leutwyler, {\it Phys. Lett.  \bf 47B} 
(1973) 365.
\item{52.} H.J.~Lipkin, in {\it Physique Nucl\'eaire}, 
Les-Houches  1968,  ed.  C.~DeWitt and V.~Gillet, Gordon and  Breach,  N.Y.
1969,  p.  585; H.J.~Lipkin, Phys. Lett. {\bf 45B} (1973) 267
\item{53.} S.C.~Frautschi, {\it Regge Poles and S-Matrix Theory}, W.A.~Benjamin Inc., 1963.
\item{54.} V.~Gribov, ZhETF {\bf 41} (1961) 1962 [SJETP {\bf 14} (1962) 1395];
V.N.~Gribov and D.V.~Volkov, ZhETF, {\bf 44} (1963) 1068.
\item{55.} H.B.~Nielsen and P.~Olesen, {\it Nucl. Phys. \bf B61} (1973) 45.
\item{56.} B.~Zumino, in {\it Renormalization and Invariance in Quantum Field 
Theory},
NATO Adv. Study Institute, Capri, 1973, Ed. R.~Caianiello, Plenum (1974)
p. 367.
\item{57.} G.~'t~Hooft, {\it Nucl. Phys.  \bf B79} (1974) 276.
\item{58.} A.M.~Polyakov, {\it JETP Lett.  \bf 20} (1974) 194. 
\item{59.} K.G.~Wilson, {\it Pys. Rev. \bf D10} (1974) 2445.
\item{60.} P.A.M.~Dirac, {\it Proc. Roy. Soc.  \bf A133} (1934) 60;
{\it  Phys. Rev. \bf 74} (1948) 817.
\item{61.} G.~'t~Hooft, ``Gauge Theories  with  Unified  Weak,  Electromagnetic  and 
   Strong Interactions", in {\it E.P.S. Int. Conf.  on  High  Energy  Physics}, 
   Palermo, 23-28 June 1975, Editrice Compositori, Bologna 1976, A.~Zichichi Ed.
\item{62.} S.~Mandelstam, {\it Phys. Lett. \bf B53} (1975) 476; {\it Phys. 
	Reports \bf 23} (1978) 245.
\item{63.} M.A.~Kramers and G.H.~Wannier, {\it Phys. Rev. \bf  60} (1941) 252, 263;
	L.~Kadanoff, {\it Nuovo Cim.  \bf  44B} (1966) 276.
\item{64.} V.~Gribov, {\it Nucl. Phys. \bf B139} (1978) 1.
\item{65.} G.~'t~Hooft, {\it Nucl. Phys. \bf B190} (1981) 455.
\item{66.} M.~Creutz, L.~Jacobs and C.~Rebbi, {\it Phys. Rev. Lett. \bf 42} (1979) 1390.
\item{67.} A.J.~van der Sijs, invited talk at the Int. RCNP Workshop on Color 
Confinement and Hadrons -- Confinement 95, March 1995, RCNP Osaka, 
Japan (hep-th/9505019)
\item{68.} T.~Massam and A. Zichichi, CERN Report, ISR Users Meeting 10-11 
June 1968 (unpublished); M.~Basile {\it et al\/}, {\it Nuovo Cimento \bf 40A}, 41 (1977)
\item{69.} H. Fukuda and Y. Miyamoto, Progr. Theor. Phys. {\bf 4} (1949) 347;
S.L. Adler, Phys. Rev. {\bf  177} (1969) 2426;  J.S. Bell and R. Jackiw, 
Nuovo Cim. {\bf  60A} (1969) 47; S.L. Adler and W.A. Bardeen, Phys.~Rev. 
{\bf  182} (1969) 1517;  W.A.  Bardeen,   Phys.~Rev. {\bf  184} (1969) 1848.
\item{70.} D.~Bollini {\it et al, Nuovo Cimento \bf 58A} (1968) 289.
\item{71.}  A.~Zichichi, {\it Proceedings of the 16th International 
Conference on "High Energy Physics"\/}, Batavia, IL, USA, 6-13 Sept. 1972 
(NAL, Batavia, 1973), Vol {\bf 1}, 145. 
\item{72.}  D.~Bollini {\it et al\/}, {\it Nuovo Cimento \bf 57A} (1968) 404.
\item{73.} J.~Kogut and L.~Susskind, {\it Phys. Rev. \bf D9} (1974) 3501, {\it ibid.
\bf D10} (1974) 3468, {\it ibid. \bf D11} (1975) 3594.
\item{74.} S.~Weinberg, {\it Phys. Rev \bf D11} (1975) 3583.
\item{75.} A.A.~Belavin, A.M.~Polyakov, A.S.~Schwartz and Y.S.~Tyupkin, 
{\it Phys. Lett. \bf 59 B} (1975) 85.
\item{76.} G.~'t~Hooft, {\it Phys. Rev. Lett.  \bf37} (1976) 8;
 {\it Phys. Rev.  \bf  D14} (1976) 3432; Err.{\it  Phys. Rev. \bf  D18} (1978) 2199.
\item{77.} R.~Jackiw, C.~Nohl and C.~Rebbi, various Workshops on QCD and 
	solitons, June--Sept. 1977; R.~Jackiw and C.~Rebbi, 
	{\it Phys. Rev. Lett. \bf 37} (1976) 172. See also: M.A.~Shifman, A.I.~Vainshtein and
	V.I.~Zakharov, {\it Nucl.~Phys. \bf B165} (1980) 45.
\item{78.} C.G.~Callan, R.~Dashen and D.~Gross, 
		{\it Phys. Lett. \bf 63B} (1976) 334.
\item{79.} R.~Crewther, {\it Phys. Lett. \bf 70 B} (1977) 359; {\it Riv. Nuovo Cim.
 \bf 2} (1979) 63; R.~Crewther, in {\it Facts and Prospects of Gauge Theories}, 
 Schladming 1978, ed. P.~Urban (Springer-Verlag 1978),; 
 {\it Acta Phys. Austriaca} Suppl {\bf XIX} (1978) 47.
\item{80.} G.~'t~Hooft, {\it Phys. Repts.  \bf 142} (1986) 357.
\item{81.} D.J.~Gross, this meeting.
\item{82.} G.~'t~Hooft, hep-th/9808113.

\bye